\documentclass{ieeeaccess}
\usepackage{cite}
\usepackage{amsmath,amssymb,amsfonts}
\usepackage{algorithmic}
\usepackage{graphicx}
\usepackage{textcomp}

\usepackage{hyperref}
\usepackage{gensymb}
\usepackage{url}

\usepackage{bm}
\makeatletter
\AtBeginDocument{\DeclareMathVersion{bold}
\SetSymbolFont{operators}{bold}{T1}{times}{b}{n}
\SetSymbolFont{NewLetters}{bold}{T1}{times}{b}{it}
\SetMathAlphabet{\mathrm}{bold}{T1}{times}{b}{n}
\SetMathAlphabet{\mathit}{bold}{T1}{times}{b}{it}
\SetMathAlphabet{\mathbf}{bold}{T1}{times}{b}{n}
\SetMathAlphabet{\mathtt}{bold}{OT1}{pcr}{b}{n}
\SetSymbolFont{symbols}{bold}{OMS}{cmsy}{b}{n}
\renewcommand\boldmath{\@nomath\boldmath\mathversion{bold}}}
\makeatother

\graphicspath{figures/}

\def\BibTeX{{\rm B\kern-.05em{\sc i\kern-.025em b}\kern-.08em
    T\kern-.1667em\lower.7ex\hbox{E}\kern-.125emX}}

\begin{document}
\history{Accepted for publication in IEEE Access\\
Published version at \url{https://ieeexplore.ieee.org/abstract/document/11015751}\\
This work is licensed under a Creative Commons Attribution 4.0 License.}
\doi{10.1109/ACCESS.2025.3573944}

\title{Wide-Angle, Multiplexed Backscatter Communications Using a Dynamic Metasurface-Backed Luneburg Lens}
\author{Samuel Kim\authorrefmark{1}, 
Tim Sleasman\authorrefmark{1}, 
Avrami Rakovsky\authorrefmark{1}, 
Ra'id Awadallah\authorrefmark{1}, 
and David B. Shrekenhamer\authorrefmark{1}}

\address[1]{Johns Hopkins University Applied Physics Laboratory, Laurel, MD 20723 USA}
\tfootnote{This project was supported by independent research and development funding from the Johns Hopkins Applied Physics Laboratory.}

\markboth
{Kim \headeretal: Wide-Angle, Multiplexed Backscatter Communications Using a Dynamic Metasurface-Backed Luneburg Lens}
{Kim \headeretal: Wide-Angle, Multiplexed Backscatter Communications Using a Dynamic Metasurface-Backed Luneburg Lens}

\corresp{Corresponding author: Samuel Kim (e-mail: samuel.kim@jhuapl.edu).}

\begin{abstract}

Backscatter communications is attractive for its low power requirements due to the lack of actively radiating components; however, commonly used devices are typically limited in range and functionality.
Here, we design and demonstrate a backscatter device consisting of a flattened Luneburg lens combined with a spatially-tunable dynamic metasurface.
Using quasi-conformal transformation optics (QCTO), we design a flattened, additively manufactured Luneburg lens that focuses incoming waves over a wide field-of-view onto its flattened focal plane.
When a reflective surface is placed at the focal plane, the flattened Luneburg lens retroreflects, enabling long-range backscatter communications over an extremely large field-of-view ($\pm30\degree$) and bandwidth.
The dynamic metasurface is designed to modulated the reflected phase across the S-band (2-4 GHz) with fine spatial control.
Thus, when combined with the flattened Luneburg lens, the device is able to modulate the retroreflected signal to achieve backscatter communications.
We experimentally demonstrate full phase control of the backscattered signal across a range of incidence angles, spatial multiplexing, and secure communications against eavesdroppers by actively suppressing or randomizing signals in unwanted directions.
The metasurface-backed Luneburg lens device offers a low-power solution for long-range wireless networks with advanced capabilities.
\end{abstract}


\titlepgskip=-21pt

\maketitle

\section{Introduction\label{sec:introduction}}

\begin{figure*}[t]
    \includegraphics[width=\textwidth]{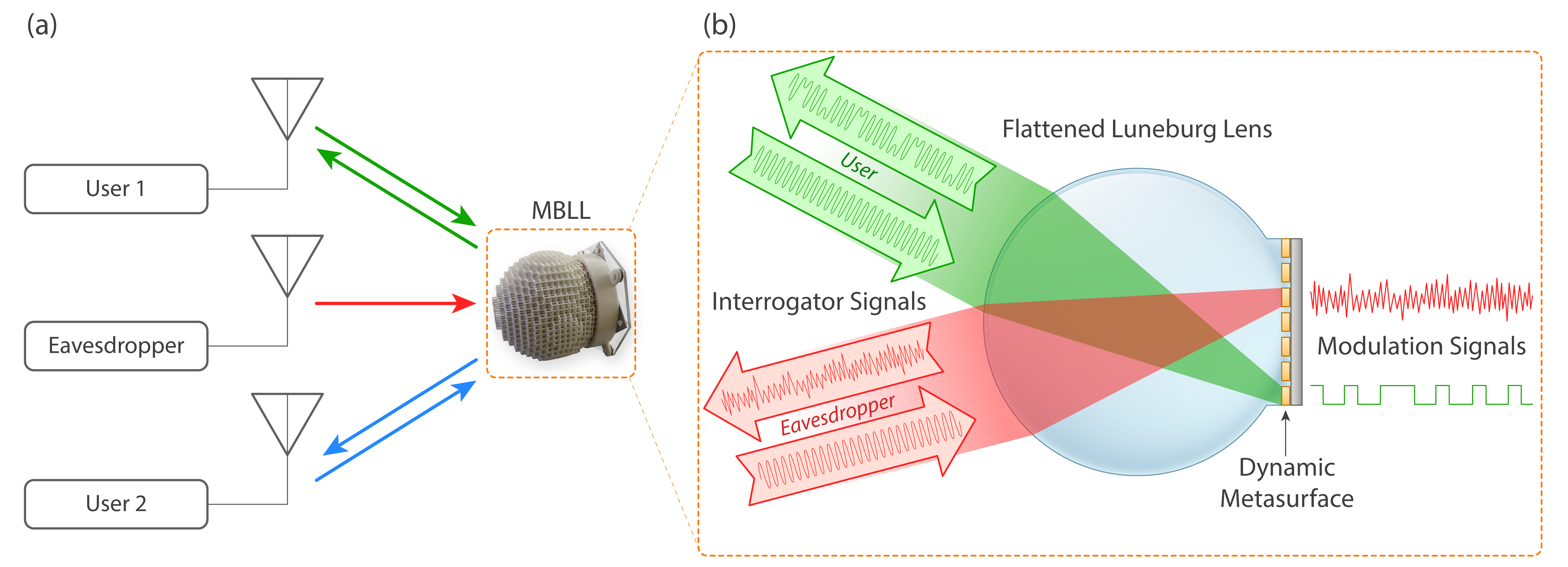}
    \caption{
    (a) Schematic of backscatter communications using the metasurface-backed Luneburg lens (MBLL).
    The MBLL is able to send separate backscatter signals simultaneously to the desired interrogators.
    The device is also able to secure the communications from the eavesdropper by either attenuating the reflected signal or randomly modulating the reflected signal with noise. 
    (b)
    The MBLL retroreflects the signal by placing the reflective metasurface at the focal plane of the flattened Luneburg lens, enabling backscatter communications. 
    The Luneburg lens focuses the incoming plane wave onto a diffraction-limited spot size on the focal plane.
    The metasurface can be modulated with spatial masks, enabling the MBLL to multiplex different signal to different interrogators. 
    Additionally, the MBLL is able to mask information from undesired directions (red signal), thus preventing eavesdropping and securing communications.}
    \label{fig:backscatter}
\end{figure*}

\PARstart{W}{ireless} communications devices, networks, and infrastructure have exploded exponentially over the past few decades, driving a variety of advancements in both hardware and software.
5G networks have become ubiquitous, taking advantage of novel technologies to provide the data rates, connectivity, and latency demanded by modern users \cite{dogra2020survey}.
The next generation of network technology such as 6G will require even more breakthroughs to provide data rates and capacity that are multiple orders of magnitude greater than before and new functionality, such as multi-sensory communications (e.g., augmented reality and telepresence) and Internet of Smart Things (IoST) \cite{alsabah20216g}.
It is estimated that nearly 16 billion devices were connected to the Internet of Things (IoT) in 2023 \cite{Vailshery_2024a}; the number of devices as well as the requirements (e.g., bandwidth, energy, range) will only continue to grow as connectivity and computer intelligence expands in applications such as smart grids, healthcare, transportation, and autonomous vehicles \cite{yalli2024internet}.

One particularly exciting technology that can address requirements for low power consumption and low cost is backscatter communications, as it does not require any radiative radio frequency (RF) components \cite{niu2019overview,liang2022backscatter}.
Instead, backscatter communications receives an incident signal from a remote RF source and modulates the reflected signal, typically by modulating the impedance of a reflective antenna.
This simplifies communication platforms by avoiding the need for active RF circuitry, thus decreasing power consumption, weight, and size requirements by potentially orders of magnitude.
Radio Frequency IDentification (RFID) devices are a common approach for backscatter communications and consist of a compact, low-frequency antenna; limited circuitry to process and modulate the signal; and in some cases, energy storage in the form of a battery or capacitor \cite{boyer2013invited}. 
While tremendously successful, many limitations prevent this technology from being used in settings with high data rates and long distances. 
For example, the fixed antenna configuration typically results in a steep trade-off between field-of-view (FOV) and gain (corresponding to range).
Additionally, RFID is subject to multipath fading in closed or cluttered environments, reducing potential data rates.

An alternative technology, reconfigurable intelligent surfaces (RIS), have attracted significant attention due to their flexibility arising from their large number of dynamically reconfigurable reflecting elements, enabling beamforming (including beam steering, multiple beams, and shaped beams) and spatial modulation \cite{nayeri2015beam,schmid2020s,liang2022backscatter}.
RISs are often implemented using reflectarrays or metasurfaces, which consist of artificially structured surfaces containing a periodic arrangement of sub-wavelength elements, where the elements can be dynamically tuned to control their electromagnetic properties.
Progress in RF dynamic metasurfaces has been substantial such that they are a variety of explorations into different tuning components \cite{saifullah2022recent} (e.g., graphene \cite{singh2020design}, field-effect transistor \cite{f2020perfect}, varactors \cite{hum2007modeling,huang2017reconfigurable,sleasman2023dual}, PIN diodes \cite{sleasman2020implementation,song2021switchable}, piezoelectric actuation \cite{rabbani2021continuous}), not to mention the nearly unlimited degrees of freedom in geometries.
The degree of control over the metasurface elements range from simple binary amplitude modulation to full grayscale control over both phase and magnitude \cite{sleasman2023dual}.

Here, we focus on the scenario where the transmitter and receiver are co-located, i.e., monostatic backscatter.
Additionally, we may wish to send different signals in different backscatter directions simultaneously or prevent eavesdropping from a listener at a different location, as shown in Fig.~\ref{fig:backscatter}. 
Reflectarrays and metasurfaces can be designed or configured to retroreflect for the monostatic backscatter scenario simply by adding a linear spatial grading to the element phases.
More generally, they can generate arbitrary beam shapes and have been demonstrated to produce multiple beams simultaneously (i.e., multibeam antennas), often relying on holographic theory to calculate the desired phase corresponding to the interference of the beams at the aperture \cite{yurduseven2017dual,qi2023steerable,qi20242d}.
Metasurfaces with a fixed feeding antenna have also been demonstrated for space- and frequency-division multiplexed wireless communications. 
However, to our knowledge, metasurfaces have not been demonstrated for multiplexed backscatter communications.
Furthermore, relying on beam shaping potentially complicates metasurface design and limits the number of possible communication bands, as each additional frequency or spatial channel negatively impacts the gain of the other channels.
On the other hand, safeguarding backscatter communications against eavesdropping has been studied for RFID \cite{f2020perfect,zhao2020safeguarding} and RIS \cite{cui2019secure,wang2023intelligent} systems.
However, these often assume a known transfer function between the transmitter, backscatter device, and receiver in order to optimize the signal reaching the desired user versus the eavesdropper.
A more elegant approach would be to integrate a separate component from the metasurface to provide the desired spatial behavior for retroreflection and enable spatial multiplexed and secure communications.

\begin{figure*}[t]
    \centering
    \includegraphics[width=0.85\textwidth]{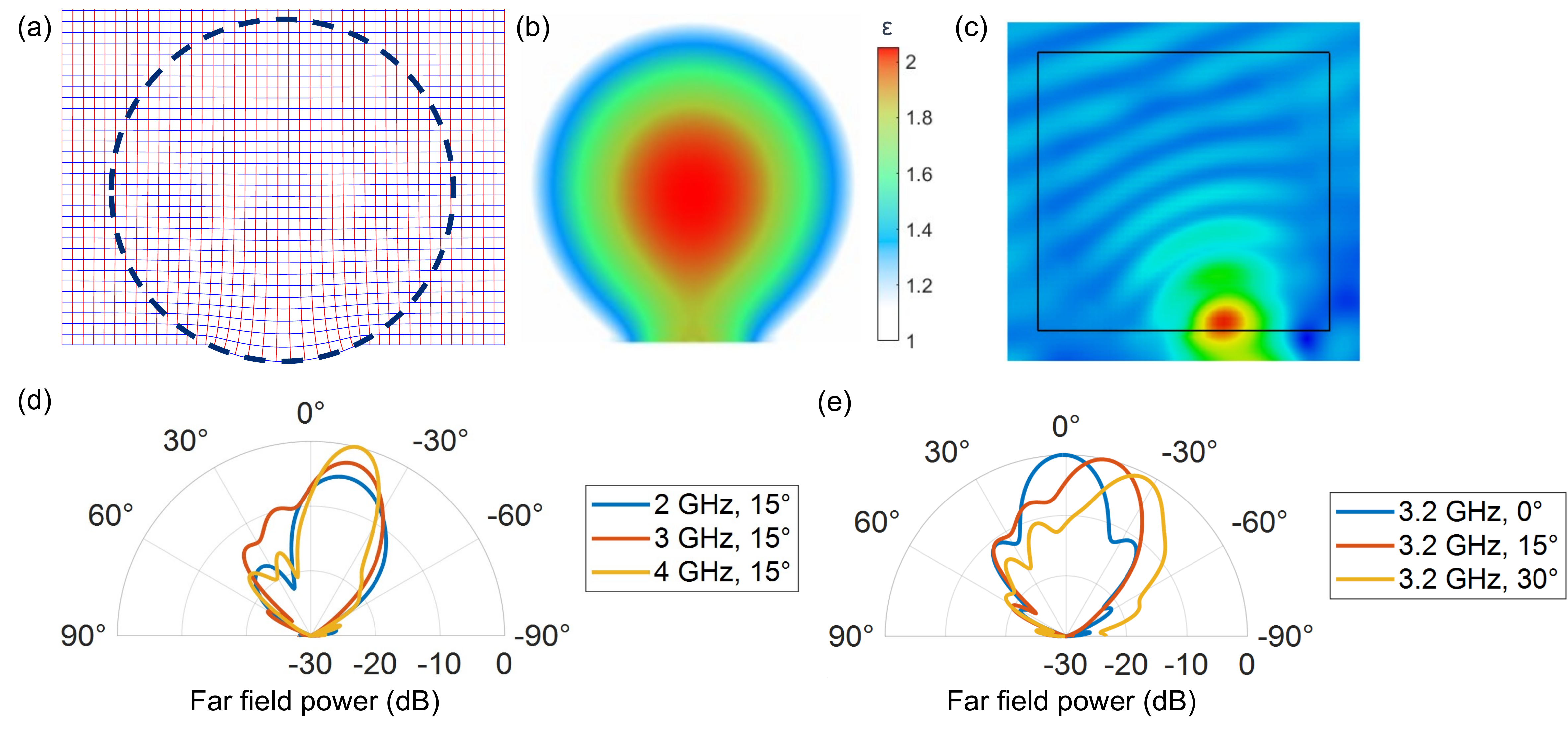}
    \caption{Flattened Luneburg lens (FLL) and simulation. (a) Coordinate transformation for flattening the Luneburg lens calculated by quasi-conformal transformation optics. Red and blue lines represent constant $x'$ and $y'$ contours in the real space plotted into the virtual space. Dotted circle represents the outline of the original Luneburg lens in the virtual space. The flattened portion represents a $50\degree$ FOV. 
    (b) Resulting permittivity of the FLL after applying fabrication-constrained approximations. 
    (c) Simulated electric field profile of a plane wave entering the FLL and focusing at the focal plane. 
    (d, e) Simulated far field pattern for the FLL with a perfect electric conductor (PEC) placed at the focal plane for 
    (d) fixed angle and various frequencies, and (e) a fixed frequency and various angles.
    Far field patterns are normalized to the peak power in the respective plots.}
    \label{fig:luneburg}
\end{figure*}

To this end, we propose a device, the metasurface-backed Luneburg lens (MBLL), for low-power, long-range backscatter communications as shown in Fig. \ref{fig:backscatter}.
When a reflective surface is placed at the Luneburg lens' focal surface, the lens retroreflects across an extremely wide field-of-view (FOV) and bandwidth. 
Thus, by modulating the reflection amplitude and/or phase, the Luneburg lens retroreflector enables long-range backscatter for a monostatic configuration.
While Luneburg lenses have been combined with modulators at the focal plane for long-range backscatter \cite{qian2023uniscatter}, the lack of spatial control over the modulator preclude the possibility of spatial multiplexing.
Thus, we combine the Luneburg lens with a reconfigurable intelligent surface (RIS), specifically a dynamic metasurface, that is able to modulate the reflection phase with fine spatial control over the surface.
Because the lens focuses signals onto a diffraction-limited spot on the metasurface, the device is able to modulate signals from different angles separately, thus enabling spatial multiplexing of backscatter communications.

Specifically, we have designed and experimentally demonstrated an MBLL consisting of a flattened Luneburg lens using quasi-conformal transformation optics (QCTO) and a varactor-loaded printed circuit board (PCB) where the elements can be independently tuned. The lens is fabricated using additive manufacturing of a metamaterial to realize the gradient index. We demonstrate (1) full backscatter phase coverage over a wide FOV ($\pm30\degree$) with improved gain and phase coverage compared to a uniform modulator, (2) binary and quadrature phase modulation of the backscatter at different angles with different signals simultaneously for multiplexed communications, and (3) modulation of the backscatter in one direction while suppressing the signal or sending random noise in another direction, preventing an eavesdropper from intercepting the data. 
\newpage

\section{Methods}
\subsection{Luneburg Lens Design}
\label{sec:lens_design}

The conventional Luneburg lenses are spherical gradient-index lenses that focus a collimated beam incident from any angle onto the lens surface.
Its spherical symmetry and effectively $360\degree$ field of view (FOV) make it attractive for wide-FOV applications.
For example, a receiver or transmitter can be placed on the focal surface to create a wide-FOV beam steering device or antenna without off-axis aberrations or multiple lenses. 
Alternatively, in the case where a reflective surface (such as a metallic layer) is instead placed on the focal surface, the device backscatters with a diffraction-limited beam directly towards the source (i.e., retroreflection) \cite{bahr20203d,kadvera2022wide}. 

However, the curved nature of the Luneburg lens focal surface makes it difficult to conformally incorporate a metasurface, which are typically fabricated using a printed circuit board (PCB) and are thus flat and rigid. 
Electronics further complicate this picture, and compatibility with a non-planar form factor restrict the metasurface design.
To overcome this, we design a flattened Luneburg lens (FLL), where a portion of the focal surface of the conventional Luneburg lens is flattened using transformation optics (TO).
TO is a computational design method that has been used to design electromagnetic and optical devices including invisibility cloaks and wave rotators.
We briefly review TO here, but more details can be found in the literature \cite{Landy2014,sun2017transformation}. 

TO aims to find the medium's electromagnetic parameters that distort the coordinate system such that the electromagnetic waves or light rays follow some desired trajectory.
In TO, we start in the virtual space with Cartesian coordinates $(x, y, z)$ that has a known medium of permittivity $\varepsilon(x, y, z)$ and permeability $\mu(x, y, z)$. 
The medium is typically chosen such that the electromagnetic behavior can be easily or even analytically described, such as an electromagnetic mode travelling down a straight waveguide, a beam focusing by a conventional refractive lens, or even a plane wave propagating in a constant medium.
The real space defined by coordinates $(x', y', z')$ represents the device we wish to realize with permittivity $\varepsilon'(x', y', z')$ and permeability $\mu'(x', y', z')$. 
Now if we desire to distort or bend the electromagnetic propagation in the virtual space to a new trajectory, we can find a coordinate transformation from $(x, y, z)$ to $(x', y', z')$ such that the device in the real space behaves as it had in the virtual space. TO gives the material parameters $\varepsilon', \mu'$ to achieve such a coordinate transformation.

The Jacobian of the coordinate transformation is defined as:
\begin{equation}
A = 
\begin{bmatrix}
\frac{\partial x'}{\partial x} & \frac{\partial x'}{\partial y} & \frac{\partial x'}{\partial z} \\
\frac{\partial y'}{\partial x} & \frac{\partial y'}{\partial y} & \frac{\partial y'}{\partial z} \\
\frac{\partial z'}{\partial x} & \frac{\partial z'}{\partial y} & \frac{\partial z'}{\partial z}
\end{bmatrix}.
\label{eq:to_jacobian}
\end{equation}
\noindent TO then gives the electromagnetic parameters in the real space as:
\begin{align}
    \varepsilon'  & = \frac{A\varepsilon A^\intercal}{\text{det}(A)} \\
    \mu' & = \frac{A\mu A^\intercal}{\text{det}(A)}.
\end{align}
The resulting parameters thus propagate electromagnetic fields the way they had in the virtual space but with the desired geometric distortion.

Note that in general, the transformed permittivity $\varepsilon'$ and permeability $\mu'$ are complex tensors that typically cannot be implemented with naturally occurring materials.
Metamaterials, which are artificially structured materials with engineered electromagnetic properties, offer a way to achieve such material profiles; however, we prefer to simplify the permittivity and permeability whenever possible such that they can be realized with isotropic dielectric metamaterials in order to achieve low loss, low frequency dispersion, and polarization insensitivity.
To this end, we use quasi-conformal TO (QCTO) to ensure that the device is able to be fabricated using non-magnetic, dielectric materials.
Quasi-conformal mappings are a type of coordinate transformation that preserve local angles, and when used for TO, result in the transformed permittivity tensor being close to scalar (i.e. isotropic) and the permeability being close to unity (i.e. non-magnetic).

Concretely, we flatten a portion of the Luneburg such that the flattened focal plane represents a $50\degree$ FOV.
The Schwarz-Christoffel mapping is used to calculate the coordinate transformation \cite{Driscoll_toolbox}, which is shown in Fig. \ref{fig:luneburg}(a).
The transformation is performed on a two-dimensional (2D) slice of the Luneburg lens, and the resulting permittivity of the flattened Luneburg lens (FLL) is shown in Fig. \ref{fig:luneburg}(b).
Applying QCTO to this device results in a slightly anisotropic permittivity, but we make the approximations $\varepsilon'_z:=\varepsilon'_r$ and $\mu':=1$.
Finally, we rotate the permittivity profile around the $z$-axis to achieve the 3D profile of the FLL

To confirm the behavior of the FLLs, full-wave electromagnetic simulations are carried out using the finite-difference time-domain (FDTD) solver in CST Microwave Studio \cite{CST}.
Fig. \ref{fig:luneburg}(c) shows the electric field on a cross-section of the lens using a plane wave source, validating the focusing of the flattened Luneburg lens.
In a separate set of simulations, the Luneburg lens is simulated with a reflective (perfect electrical conductor) backing at the focal plane to confirm the retroreflective behavior of the lens across multiple frequencies and incidence angles, as shown in Fig. \ref{fig:luneburg}(d,e).
All of the farfield patterns display a main lobe at the incident angle, confirming retroreflection.
The $15\degree$ incidence angle simulations have main lobes with 3dB beamwidths of $21.7\degree$, $16.5\degree$, and $11.6\degree$ at 2, 3, and 4 GHz, respectively. 
The side lobes levels are -24~dB, -14~dB, and -32~dB for the 3 respective frequencies, which are partially due to finite aperture of the lens and the non-Gaussian source.
While the size of the main lobes do not vary much with incidence angle, the side lobe levels tend to increase with increasing incidence angle.

Finally, we note that using any lens or even any transformation to the Luneburg lens does not guarantee retroreflective behavior. In Appendix \ref{app:hemispherical}, we look at a hemispherical Luneburg lens proposed in Ref. \cite{xu2022hemispherical} which does not achieve retroreflection in simulation.

\begin{figure*}[tb]
    \includegraphics[trim={0 15pt 0 0},clip,width=\textwidth]{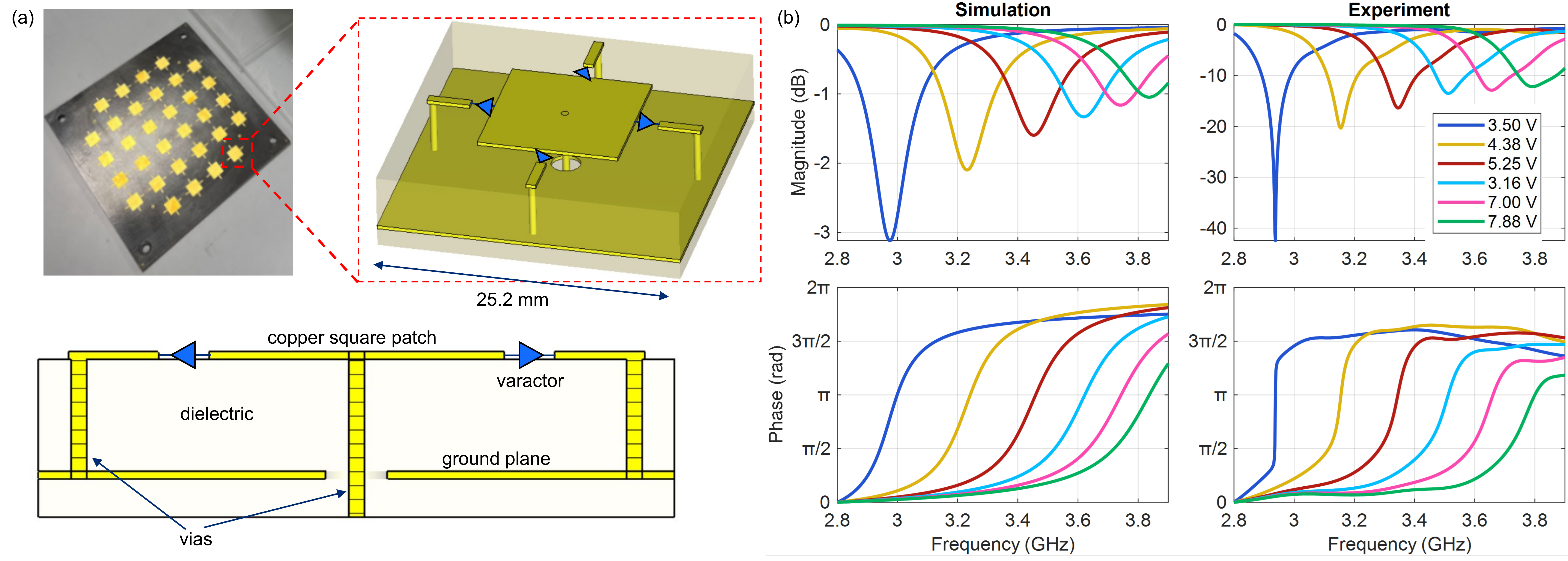}
    \caption{Dynamic metasurface for modulating phase of the reflected signal. (a) Image of the fabricated metasurface and schematic of a single unit cell (schematic not to scale). The metasurface consists of a $6\times6$ array of unit cell excluding the corners. The unit cell consists of a square metal patch over a metal-backed ground plane. The patch is connected to varactors (blue triangles) which are connected by through-hole vias to the ground plane. (b) Simulated and experimental results of the reflection amplitude and phase at normal incidence. Experimental results are normalized to the maximum magnitude measured for each voltage setting. Phase is manually corrected to account for the propagation distance, and is normalized to 0 phase at 2.8 GHz to facilitate visualization. }
    \label{fig:metasurface}
\end{figure*}

\subsection{Luneburg Lens Fabrication}

The FLL has a gradient-index profile and is thus difficult to implement with naturally occurring materials; we turn to dielectric metamaterials to serve as an effective medium.
Metamaterials, which are artificially structured materials to achieve properties not available in bulk materials, have been used to realize Luneburg lenses in both 2D and 3D \cite{larimore2018additive,kim2019luneburg,biswas2019realization,guo20213d}. 
Here we use a metamaterial unit cell consisting of a rectangular strut lattice where the strut width may vary in each unit cell to achieve the desired permittivity.
The Maxwell-Garnett approximation is used to calculate the width of the struts.
Note that as a matter of terminology, the FLL is physically realized using a 3D meta\textit{material}, which differs from the dynamic meta\textit{surface} that we place on the focal surface of the Luneburg lens.

The resulting metamaterial-based lens is realized with additive manufacturing. 
We use a powder-bed printer (EOS P395) using selective laser sintering (SLS) to manufacture the FLL with EOS PA3200 silica-loaded nylon, which has a permittivity of 2.8.
The strut widths range from 1 mm to 3.095 mm and the metamaterial unit cell size is set to 5 mm.
This allow the metamaterial to achieve a continuous range of permittivities from 1.12 to 2.02.
The diameter of the Luneburg lens before applying QCTO is 24 cm.
After flattening and truncating the lens to regions with permittivities that are achievable by the metamaterial, the size of the realized lens is 21.59 cm in diameter and 21.71 cm in height.

\subsection{Metasurface Design}

The metasurface unit cell consists of a square metal patch on a printed circuit board (PCB) loaded with MAV-000120-1411 varactors, as shown in Fig. \ref{fig:metasurface}(a). 
The unit cell design is targeted for S-band operation (2--4 GHz) and wide-angle operation so that the focusing of the FLL onto the metasurface does not distort the phase response.
The square patches measure 11.7 mm across while the center-to-center distance is 25.2 mm. 
The center of the patch is connected to a control voltage on the back of the metasurface through a via, where the control voltage can vary between unit cells. 
The varactors connect between the patch and the ground plane through a via. 

The unit cells are arrayed in a 6x6 grid with the corners removed as shown in Fig. \ref{fig:metasurface}(a), which is sufficient to cover the flattened focal plane of the FLL.
A controller (MCC USB-3114) with 16 distinct voltage levels is used to tune the elements. 
The central 16 elements in the array each have a distinct voltage, and the boundary elements are connected to their nearest neighbor.
This allows for independent grayscale tuning of the patches, which results in control over the reflection phase (and some magnitude variation based on the element’s resonance).

A comparison of the experimental and simulation results of the metasurface are shown in Fig. \ref{fig:metasurface}(b-e).
The simulations are carried out in the Frequency Domain (FD) solver in CST Microwave Studio \cite{CST} and consist of a single unit cell with periodic boundary conditions.
The experimental setup is described in Section \ref{sec:results}.
The frequency of the resonances match closely between the experimental and simulations results.
However, the reflection amplitude dips of the experimental results at resonance are markedly larger than those of the simulation results, signifying additional losses that are not accounted for.
We suspect that this extra loss is due to the electroless nickel immersion gold (ENIG) surface finish that was used to coat the copper, which was not modeled in the results shown in Fig. \ref{fig:metasurface}(d-e).
Indeed, when a nickel coating is added to the simulations, the loss increases to a similar magnitude as the experimental results.
Future work will use lower loss surface finishes.
We note that the same metasurface is used in the results going forward and will similarly show the increased loss.
However, we still compare with the low-loss simulation results to demonstrate the potential performance of the device.

\begin{figure*}[t]
    \includegraphics[width=\linewidth]{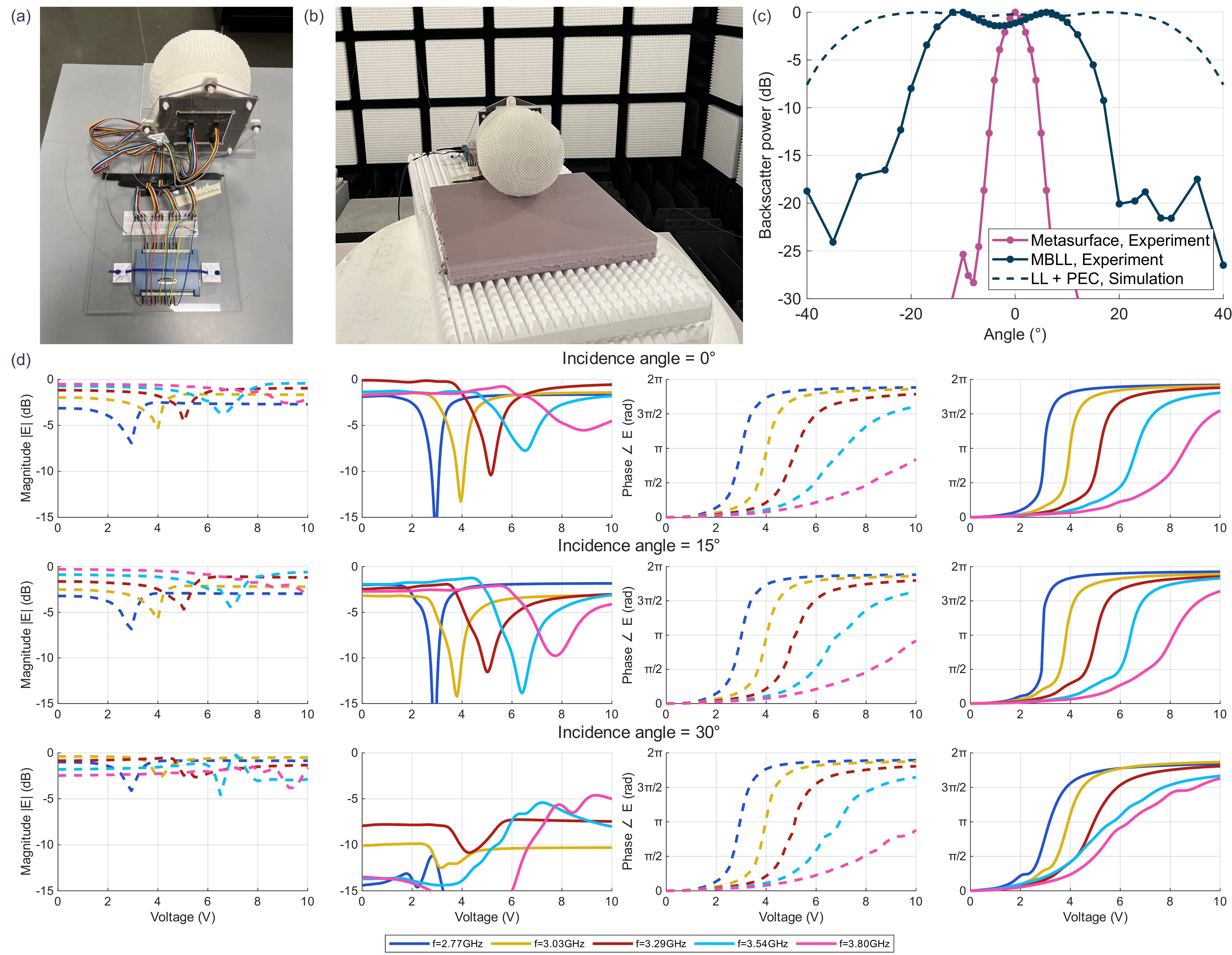}
    \caption{
    (a) Back view of the MBLL and the voltage controller. (b) Experimental setup of the MBLL in the anechoic chamber. 
    (c) Backscatter (retroreflection) measurements for the MBLL as well as a metal plate for comparison. Simulation of the MBLL uses a perfect electric conductor (PEC) plate at the FLL focal plane to reduce computational complexity. 
    (d) MBLL backscatter field magnitude and phase as a function of metasurface tuning, which is uniform across the metasurface. 
    The metasurface achieves almost a full $2\pi$ phase control up to $30\degree$ incidence angle, although the magnitude starts to degrade above $15\degree$. 
    Dotted lines refer to simulations and solid lines refer to experimental data.
    }
    \label{fig:lens_pic}
\end{figure*}

\subsection{Metasurface-backed Luneburg Lens}

The dynamic metasurface is attached to the flattened focal plane of the FLL as shown in Fig  \ref{fig:backscatter} and Fig. \ref{fig:lens_pic}(a).
A solid ring with screw holes has been added to the perimeter of the FLL base for structural support and to allow attachment of the metasurface using screws.
The dynamic metasurface modulates the signal retroreflected by the FLL, and so the combination of the two components, the MBLL, enables high-gain, wide-FOV backscatter communications.
Because the FLL focuses incoming radiation to a diffraction-limited spot, the spatial variation of the metasurface allows for signals from different angles to be modulated independently, enabling spatial multiplexing of communications. 
Thus, the MBLL features the most useful aspects of its respective components. 

We note that the simulation performance of the MBLL improves in terms of loss and angular dispersion when a small gradient-index impedance-matching layer of 2~mm thickness is inserted at the focal plane of the Luneburg lens.
This is likely because the FLL has a non-unity permittivity at its focal plane (where the metasurface lies), which may impact the metasurface behavior through dielectric loading since the metasurface was designed for a surrounding medium of unity index (i.e., air).
Additionally, the impedance-matching layer does not need to be thick (e.g., on the order of the wavelength) since the surface modes of the metasurface are largely confined to the metal and dielectric and decay quickly away from the metasurface.
In practice, because this impedance-matching layer is smaller than the unit cell size of the metamaterial used to construct the Luneburg lens, we simply add a standoff distance of 1~mm between the lens and the metasurface.
Any remaining perturbation of the reflection resulting from the varying permittivity of the FLL can be compensated for by the spatially-tunable dynamic metasurface.

Also note that this thin impedance-matching layer serves a different function than an anti-reflective coating, which are typically on the order of the wavelength to minimize boundary reflections, and have been applied to the flattened Luneburg lenses \cite{biswas2020high,kadvera2022wide}.
However, we have found the decrease in reflection magnitude to be fairly minimal, likely due to the relatively small perturbation in permittivity when applying transformation optics in our design.
Future work expanding the FOV of the system may require a more significant coordinate transformation, and thus a larger anti-reflective coating to minimize losses.

\section{Results}
\label{sec:results}

Measurements are taken with a vector network analyzer (VNA) in an anechoic chamber, where a single directional horn is used as the source and receiver in a monostatic configuration. 
The MBLL is placed on its side on a turntable to assess angular performance, such that the turntable rotates the lens along its elevation angle relative to the horn.
This setup is shown in Fig. \ref{fig:lens_pic}(b).
Another copy of the metasurface without the FLL is also fabricated and measured to separately characterize the behavior of the FLL and the metasurface.
Undesired signals are time-gated to isolate the response from the MBLL. 

Fig. \ref{fig:lens_pic}(c) shows the backscatter of a rectangular metal plate which serves as a reference, and the MBLL where the metasurface is tuned off-resonance such that it acts as a conductive metal sheet. 
We confirm that the FLL provides retroreflection across a wide field-of-view (FOV), matching the simulated results. 
The measured 3-dB FOV is $36\degree$, which is lower than the simulated FOV of over $60\degree$; this may be due to the additional support structures that were added to the region around the focal plane of the FLL to attach it to the metasurface.
Future work can optimize the support structures so that it does not interfere with the performance of the lens.
Furthermore, we note that the backscatter has a slight dip at $0\degree$ which we also confirm in simulation at certain frequencies; 
we conjecture that this may be due to the higher index of the Luneburg lens at the center of the focal plane which may result in a slight impedance mismatch.
This can be addressed through an impedance-matching layer at the focal plane, which have been demonstrated for Luneburg lenses \cite{kadvera2022wide}.

\begin{figure}[t]
    \includegraphics[width=0.95\columnwidth]{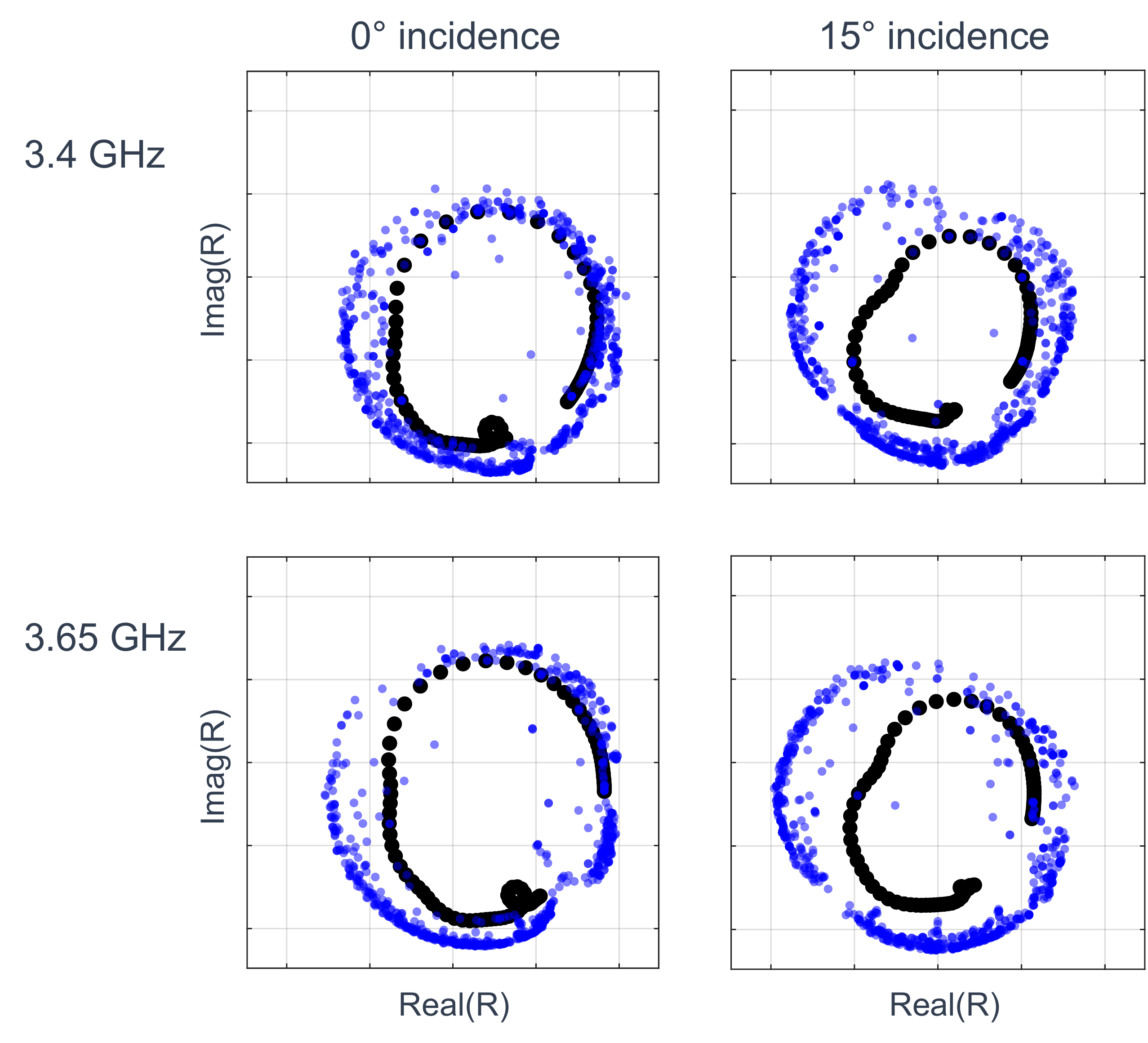}
    \caption{Constellation diagrams for MBLL backscatter at different frequencies and incidence angles. Black points represent the spatially uniform masks and the blue points represent the optimized masks to maximize return signal and fill out the phase. Each combination of frequencies and incidence angles were optimized separately. Axes are arbitrarily normalized.}
    \label{fig:ga}
\end{figure}

\subsection{Full phase control}
\label{sec:full_phase}

By tuning the varactor voltage across the metasurface, we can modulate the backscattered phase across a wide FOV and range of frequencies, as shown in Fig. \ref{fig:lens_pic}(d). 
Here, the varactor voltage is uniform across the metasurface unit cells.
The simulations, which are carried out using the CST Microwave Studio FDTD solver, consist of a 2.5-dimensional slice of the MBLL rather than the full 3-dimensional MBLL due to computational expense arising from the need for fine meshing of the metasurface.
The experimental data is also represented as a constellation plot in Fig. \ref{fig:ga} (black points).
The backscattered signal for the MBLL achieves a phase coverage of $>300\degree$ for an angular swath exceeding $\pm 30\degree$, which is consistent with the reflection profile of the bare metasurface without the FLL in Fig. \ref{fig:metasurface}(b-e).
Thus, the FLL does not negatively impact the performance of the metasurface.

To fully exploit the capabilities of the MBLL, we can apply different voltages to each unit cell of the metasurface to achieve a spatially-varying phase response across the metasurface---which we will refer to as spatial masks---and assess how the masks change the reflected backscatter (both magnitude and phase). 
While we have shown that sweeping a uniform voltage of the varactors across the metasurface can achieve a $300\degree$ phase coverage in the backscatter, it is possible that the phase range can be improved further using non-uniform voltage masks due to the interaction between the FLL and the metasurface as well as the finite edge effects of the metasurface.
Thus in the first experiment, we optimize the non-uniform spatial masks to the metasurface to optimize the backscatter for full magnitude and phase control.
We apply a custom optimization algorithm based on multi-objective genetic algorithms to maximize the magnitude of the signal for a set of discretely sampled phases from $0\degree$ to $360\degree$.
The results are shown in Fig. \ref{fig:ga} and more details on the optimization can be found in Appendix \ref{app:optimization}. 
The optimized spatial masks are indeed able to not only find patterns with greater magnitude than the uniform patterns, but are also able to fill in the gap and achieve a full $360\degree$ phase response.

\subsection{Multiplexed Communications} \label{sec:multiplex}

\begin{figure}[t]
\centering
\includegraphics[width=0.85\columnwidth]{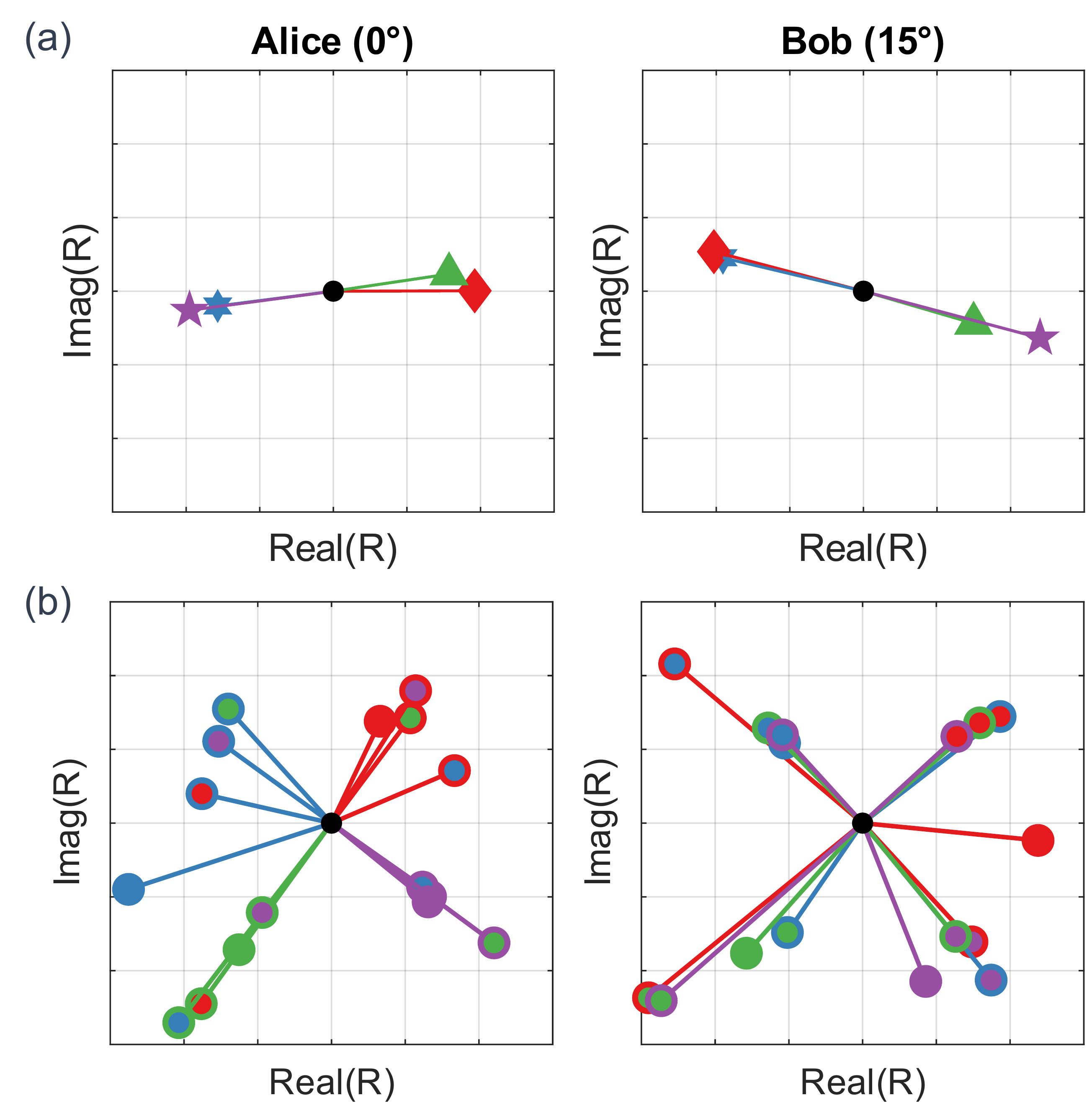}
\caption{Demonstration of multiplexed communications to two different users---Alice (at $0\degree$) and Bob (at $15\degree$).
(a) Constellation diagram of 4 spatial masks that is able to achieve every combination of 0 and $\pi$ phases (corresponding to ``0'' and ``1'' bits for BPSK) to Alice and Bob.
Each color/marker shape represents a unique spatial mask for the metasurface. 
Mask search is performed at 3.76 GHz.
(b) Constellation diagram of 16 spatial masks that is able to achieve 4 different phases to Alice and Bob, thus enabling multiplexed QPSK.
Each combination of marker face colors and line colors represents a unique spatial mask for the metasurface (e.g., blue line with red marker represents the same mask for Alice and Bob).
Mask search is performed at 3.8 GHz.
}
\label{fig:multiplex}
\end{figure}

The MBLL enables spatial multiplexing of backscatter signals since the FLL focuses electromagnetic waves incident from different angles onto different spatial locations on the metasurface, which can be modulated separately.

While optimization algorithms will likely deliver ideal results, here we relegate our analysis to a simpler subset of masks of 5000 randomly sampled metasurface spatial patterns due to the limitations of the experimental setup.
The reflection of the MBLL with the 5000 spatial masks is measured at $0\degree$ incidence (which we name Alice) and $15\degree$ incidence (which we name Bob).
We wish to find spatial masks such that Alice and Bob receive different bit streams from the MBLL, similar to the case in Ref. \cite{xu2024two}.
Importantly, note that communication is not taking place between Alice and Bob, but rather that the MBLL communicates back separate information to each users when simultaneously interrogated by the two users.

In a binary phase-shift keying (BPSK) implementation each user needs to receive two phases---0 and $\pi$ corresponding to bits 0 and 1. 
Since there are two users and two phases, we aim to find four masks which can pass the different data permutations. 
We perform a simple search over the 5000 masks to four masks that minimizes the following loss function:
\begin{multline}
\mathcal{L} = |\alpha_A(\mathbf{x}_1) - \phi_{A,0}| + |\alpha_B(\mathbf{x}_1) - \phi_{B,0}| + \\
|\alpha_A(\mathbf{x}_2) - \phi_{A,0}| + |\alpha_B(\mathbf{x}_2) - \phi_{B,1}| + \\
|\alpha_A(\mathbf{x}_3) - \phi_{A,1}| + |\alpha_B(\mathbf{x}_3) - \phi_{B,0}| + \\
|\alpha_A(\mathbf{x}_4) - \phi_{A,1}| + |\alpha_B(\mathbf{x}_4) - \phi_{B,1}|
\label{eq:multiplex_loss}
\end{multline}
where $\mathbf{x}_i$ represents the four different spatial masks, $\alpha$ represents the backscatter phase, $\phi_{\{0,1\}}$ is the desired phase for bits 0 and 1, and the subscripts $A$ and $B$ represent Alice and Bob, respectively. 
In BPSK, the bits should be $\pi$ phase apart such that $\phi_1=\phi_0+\pi$. 
Note that in general, $\phi_A$ does not necessarily equal $\phi_B$ since these are separate communication channels. 
Additionally, the desired phase $\phi_0$ is also a free parameter.

A constellation diagram of a candidate set of masks at 3.76 GHz is shown in Fig. \ref{fig:multiplex}(a) where the different marker colors or shapes correspond to different spatial masks, $x_i$.
Each user clearly has bits that are $\pi$ phase apart, and there exists a mask for every combination of bit pairs to Alice and Bob.
For example, if the green triangle represents $(A=0,B=0)$, then the red diamond represents $(A=0,B=1)$ and the purple triangle represents $(A=1,B=0)$.
Note that the relative phase of Alice and Bob is arbitrary.
This result demonstrates that the MBLL is able to communicate two separate bit streams for the two users.

Extending upon this, we also demonstrate the potential for quadrature phase-shift keying (QPSK) in which the communications relies on four points equispaced around a circle in the constellation diagram.
Each point in the constellation diagram encodes two bits, thus doubling the data rate of the signal for a constant bandwidth.
Using a similar optimization scheme as before over the 5000 random masks, we find 16 different masks that can communicate every possible pair of constellation points to Alice and Bob.
The resulting constellation plots are shown in Fig. \ref{fig:multiplex}(b).
Here, each pair of marker face color and line color represents a unique mask.
For example, the red line with a green marker face represents the same mask for Alice and Bob.
We see that all of the 4 points with a red line represent the same bit for Alice, and each of these 4 points corresponds to a unique pair of bits for Bob.
Similarly, the 4 points with green face color represents every possible pair of bits for Alice, and yet corresponds to the same bit pair for Bob.
Thus, these 16 spatial masks can be used to encode QPSK for Alice and Bob simultaneously.

Again, the results here are based on randomly sampled metasurface masks, but more sophisticated sampling or optimization methods may be applied to further improve the results.
For example, the MBLL can be likely be extended to higher-order phase modulation techniques for even higher data rates, or multiplex to more than two users.
In a simplified model of the MBLL, we can imagine modulating the unit cells only at the focal spot, and thus enabling spatial multiplexing with the number of channels on the order of (metasurface area) $/$ (focal spot size).
As we have seen from the results in Section \ref{sec:full_phase}, this is indeed a simplified picture as the performance can be improved through the use of spatially varying metasurface masks.

\begin{figure*}[t]
\centering
\includegraphics[width=\textwidth]{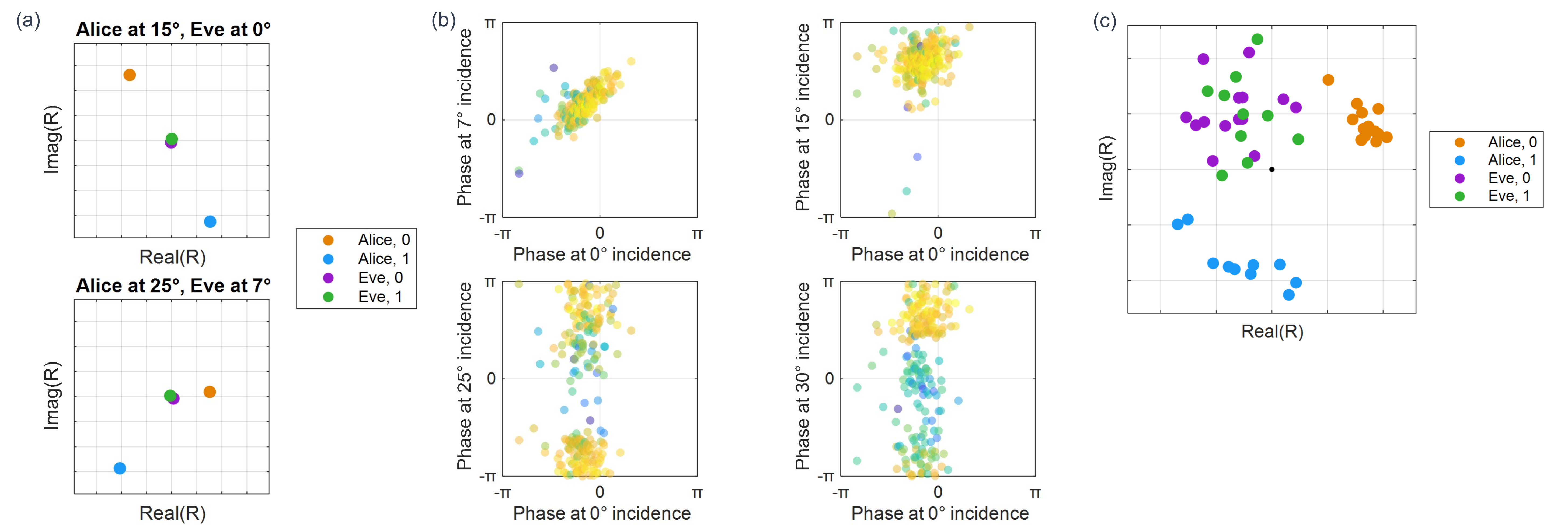}
\caption{Demonstration of the MBLL for secure communications (to Alice) in the presence of an eavesdropper (Eve). 
(a) Constellation diagram corresponding to a pair of masks that maximize the phase variation for Alice while minimizing the backscatter signal for Eve. 
In the first configuration, Alice and Eve are at $15\degree$ and $0\degree$ incidence, respectively. The masks are found from searching through 5000 random masks at 3.75 GHz. The smaller of Alice’s symbols and larger of Eve’s symbols differ by 32.2dB. 
In the first configuration, Alice and Eve are at $20\degree$ and $7\degree$ incidence, respectively. The masks are found from searching through 500 random masks at 3.67 GHz. The smaller of Alice’s symbols and larger of Eve’s symbols differ by 32.2dB. 
(b) Backscatter phase from 500 random spatial metasurface patterns at various angles. $x$-axis corresponds to Alice and $y$-axis corresponds to Eve, allowing us to see the correlation between the two interrogators. Points are colored according to signal magnitude at Eve.
(c) Constellation diagram corresponding to a set of masks that can be used for secure communications, where Alice is at $0\degree$ and Eve is at $15\degree$.
}
\label{fig:eve}
\end{figure*}

\subsection{Preventing eavesdropping}

Finally, we consider the scenario where an eavesdropper (Eve) attempts to capture the communication between Alice and the MBLL.
Specifically, suppose Alice is interrogating the MBLL from some incidence angle and an eavesdropper is attempting to capture the communication between the MBLL and Alice from a different angular position by detecting the signal from the MBLL towards Eve.
We consider two possible methods for eavesdropping: passive and active.

In passive eavesdropping, Eve is not interrogating the MBLL, but is rather attempting to capture stray reflections from Alice interrogating the MBLL.
Although the Luneburg lens ideally provides retroreflection towards the interrogator, there may be sidelobes away from the interrogator due to the finite aperture of the lens.
As discussed in Section \ref{sec:lens_design}, the simulated sidelobes for Alice at $15\degree$ incidence are generally below $-14$~dB and reach as low as $-32$~dB. 
If the background noise is greater than those sidelobe levels, the eavesdropper will not be able to differentiate the signal from the noise.
The sidelobe levels can likely be further suppressed by optimizing the metasurface spatial masks using different objective functions, which we leave for future work.

In active eavesdropping, Eve is interrogating the MBLL simultaneously with Alice.
Since the metasurface is being reconfigured to modulate the backscatter signal to Alice, the reconfiguration may also imprint a modulation on the backscatter to Eve.
One possible strategy to secure communications is to find spatial masks that maximize the signal for Alice while minimizing the signal for Eve such that the signal is below the noise threshold.
For this experiment, we sample random spatial patterns for the metasurface and measure the backscatter at various angles.
In the first configuration, we suppose that Alice is at $15\degree$ incidence angle and Eve is at $0\degree$ incidence, and search for a pair of masks from 5000 random masks at 3.75 GHz that maximize the signal for Alice---which in the case of BPSK would be two masks with $\pi$ phase difference---while maximizing the power ratio of Alice's backscatter to Eve's backscatter signal.
In the second configuration, we suppose that Alice is at $25\degree$ incidence and Eve is at $7\deg$ incidence, and again search for a pair of masks, this time from 500 random masks sampled at 3.67 GHz.
A constellation diagram for the backscatter of one such pair of masks in each configuration is shown in Fig.~\ref{fig:eve}(a).
The two masks are sufficiently separated in phase space for Alice to implement a code such as BPSK.
Additionally, the power of the greatest signal for Eve (purple dot) is 32.2~dB and 24.6~dB lower than the power of the smallest signal for Alice (orange dot) in the two configurations, respectively
Thus, if the MBLL is configured such that the interrogator signal is less than this signal difference above the noise floor, then Eve will not be able to eavesdrop on the communication with Alice.

Alternatively, it is conceivable that with a sufficient amount of power in the transmitter and/or a sufficiently powerful receiver, Eve could detect the remnant phase variations in the backscatter signal which would be correlated with Alice's phase variations, thus successfully capturing the communications.
To counter this, we also consider an alternate strategy of finding multiple masks for each symbol such that the masks present a large enough variation for Eve so as to hide the original symbol.
For communicating a particular symbol to Alice, the mask can be randomly chosen from the set of suitable candidates.
The backscatter signal at Eve, however, should appear uncorrelated with the backscattered signal at Alice.
Thus, even though the signal for Eve may be above the noise floor, the symbols will be random and appear as noise, thus securing communications.

For this experiment, we sample 500 random spatial patterns for the metasurface and measure the backscatter at several angles: $0\degree$, $7\degree$, $15\degree$, $25\degree$, and $30\degree$.
We assume that Alice is at $0\degree$ incidence angle and that Eve is at any other angle.
Fig. \ref{fig:eve}(b) plots the measured backscatter phase, where the $x$-axis and $y$-axis corresponds to the backscatter phase for Alice and Eve, respectively.
We see a loose correlation between the phase at $0\degree$ incidence and $7\degree$ incidence which is unsurprising due to the strongly overlapping focal spots on the FLL focal plane.
Notably, the correlation falls off above $15\degree$ incidence, suggesting that we can mask Alice's message from a sufficiently distant Eve.
In particular, we can choose a set of masks that are approximately $\pi$ phase apart for Alice but where the phase is ambiguous for Eve, as shown in Fig.~\ref{fig:eve}(c).
Thus, if we want to send a a ``0'' or a ``1'' to Alice, we can choose from any of the masks corresponding to an orange or blue point, respectively.
Because distributions of phase are similar to Eve regardless of the bit to Alice, the message is masked from Eve.

Note that in both experiments, the masks have not been optimized due to experimental complexity.
It is likely that these results can be further optimized to maximize the signal to Alice while minimizing the signal for Eve.
Future experiments can also investigate the possibility of multiple interrogators or eavesdroppers across locations or frequencies.

\section{Conclusion and Future Work}

We have designed and experimentally demonstrated a MBLL for wide-FOV and long-range retroreflection which is capable of spatially-multiplexed and secure backscatter communications. 
The use of the FLL to achieve retroreflection alleviates the metasurface design and control constraints, enabling low-loss backscatter across a broad range of frequencies and angles.
Applying spatial masks to the metasurface improves gain, enables communications with two users simultaneously, and secures communications from an eavesdropper.

The MBLL has potential for a wide variety of applications.
Reconfigurable intelligent surfaces (RISs) and metasurfaces have long been proposed for backscatter communications, with applications in wireless communications and internet of things (IoT) \cite{niu2019overview,liang2022backscatter}.
The retroreflective behavior of the MBLL as well as the ability to secure communications can reduce interference in cluttered environments such as indoor, warehouse, or urban settings where multipath losses can degrade backscatter channels. 
The high-gain of the MBLL enables backscatter communications over long distances, which can enable applications such as vehicle identification for intelligent traffic management \cite{qian2023uniscatter}. 
The device may be attached to a sensor to share data without consuming significant power, which can have applications such as monitoring plants in agricultural settings  or monitoring infrastructure for smart cities. 
Numerous other possibilities exist for augmenting the MBLL and there remain many applications that motivate continued development of the MBLL.

While we choose a simple form for the metasurface unit cell which achieves full $2\pi$ phase coverage, the FLL can easily be combined with other types of metasurfaces with different functionality.
For example, Ref. \cite{sleasman2023dual} demonstrate a metasurface for independent control over both reflection amplitude and phase.
This can enable even higher bitrate communication protocols compared to phase shift keying (PSK) demonstrated here, such as quadrature amplitude modulation (QAM).
Other types of multiplexing such as frequency, polarization, or orbital angular momentum (OAM) can potentially increase channel capacity even further.

Many of the results presented here using spatial masks for the metasurface rely on either simple optimization schemes (genetic algorithms) or random sampling, as more advanced optimization is outside the scope of this work.
The simple approaches taken here already demonstrate impressive results, and there is tremendous potential in applying different optimization approaches to achieve improved behavior from the MBLL that would not be possible otherwise.
For example, channel estimation aims to model and optimize the communication channel characteristics between a transmitter and receiver, and has been applied to RISs \cite{swindlehurst2022channel,zhang2023channel}.
These techniques can possibly be used to efficiently optimize the metasurface configuration without extensive global optimization algorithms, although they often assume perfect control over the phase response.
Additionally, machine learning has been applied to optimizing beamforming in RISs \cite{faisal2022machine}.

\appendices

\section{\break Optimization}
\label{app:optimization}

To find a set of spatial masks that can maximize the backscatter magnitude while simultaneously achieving a full range of phase response, we use multi-objective optimization.
Rather than minimizing or maximizing a single metric as is done in conventional optimization, multi-objective optimization aims to find a set of solutions called the \textit{Pareto front} that balance the trade-off between the multiple objectives.
In particular, we use the multi-objective genetic algorithm included in MATLAB \cite{datta2007multi, MATLAB}. 
The two objectives are:
\begin{align}
    \mathcal{L}_1 & =|S| \\
    \mathcal{L}_2 & = |(\angle{S}) - \phi_0| / 10\degree
\end{align}
where $\phi_0$ is the desired angle, and the division by $10\degree$ in the second objective is to normalize the two objectives to approximately the same magnitude for purposes of setting meaningful convergence criteria. 
The genetic algorithm is placed inside a loop that iterates over 16 equally spaced phase targets from 0 to $2\pi$ radians.
The collective Pareto fronts from all previous optimizations are used as part of the initial population for the next optimization to improve convergence, where the Pareto front is re-calculated for the new objective and a clustering algorithm (k-means) is used to down-select the candidates.

Note that a constrained single-objective optimization approach where the phase is set as a constraint may be more meaningful in the case where we wish to optimize for a single angle. However, in the case where we wish to optimize for a collection of angles, the multi-objective optimization approach works reasonably well in practice. 

Finally, note that this is a simple optimization scheme that ignores the magnitude of the backscatter field.
We preprocess the dataset by filtering out the masks with a magnitude below a threshold.
In practice, this works to find reasonable solutions; alternatively, more sophisticated methods such as multi-objective optimization or an objective more closely tied to the modulation scheme can be used to maximize signal-to-noise ratio and data rates.

\section{\break Hemispherical Luneburg Lens}
\label{app:hemispherical}

\begin{figure*}[htbp]
\includegraphics[width=0.8\textwidth]{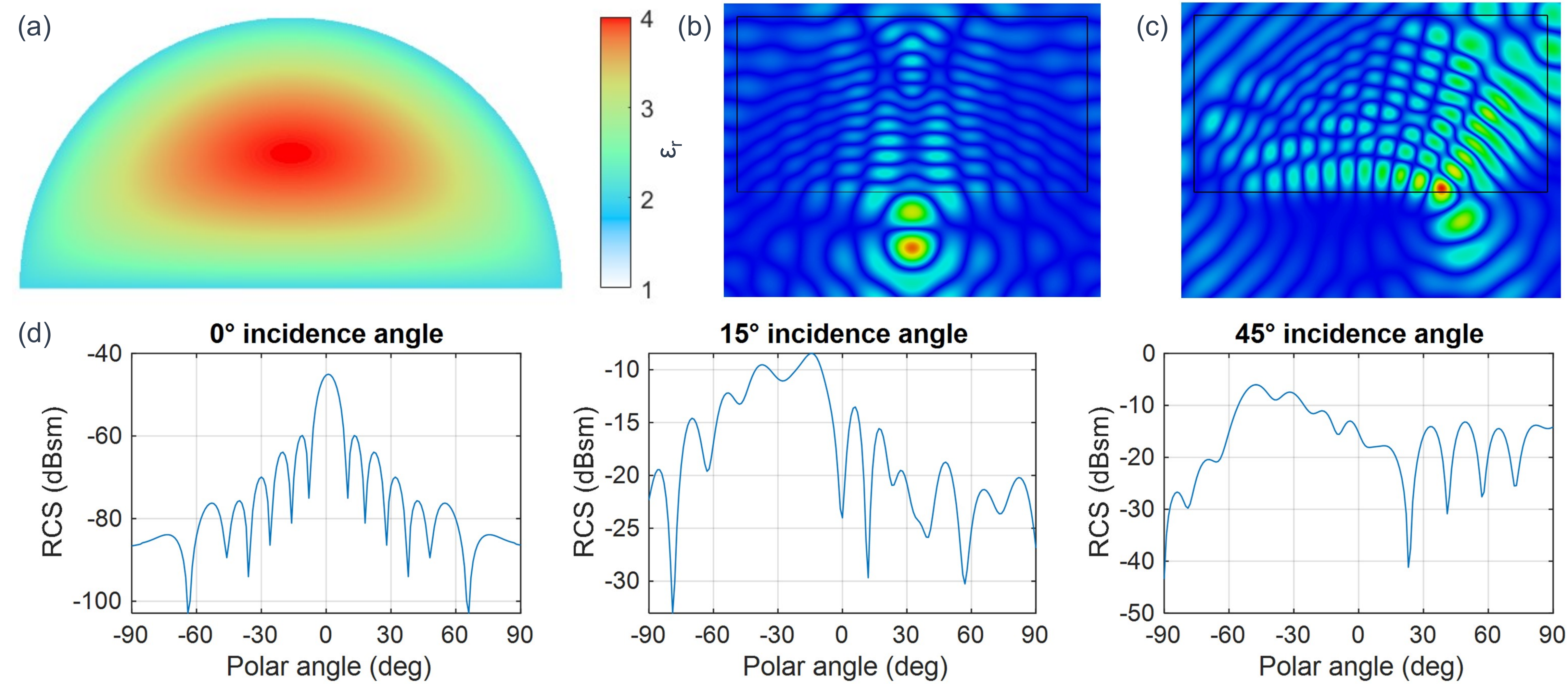}
\centering
\caption{Simulations of a hemispherical Luneburg lens flattened using the transformation optics mapping proposed in Ref. \cite{xu2022hemispherical}. (a) Relative permittivity profile. (b,c) Electric field magnitude calculated using full-wave simulations for plane waves at (b) $0\degree$ and (c) $45\degree$. Black rectangle outlines the region of the hemispherical lens. (d) Far field radar cross section (RCS) of the lens when a reflective surface (PEC) is placed at the flattened surface of the lens.  }
\label{fig:hemispherical}
\end{figure*}

The coordinate mapping used in transformation optics is not unique.
While the quasi-conformal mapping is one of the more commonly used mappings due to its ability to be realized using isotropic, non-magnetic materials, other mappings can be used with various tradeoffs.
For example, Xu et al. propose a hemispherical Luneburg lens using an analytical coordinate mapping in order to integrate the lens with an array of feed antennas \cite{xu2022hemispherical}.
However, we note that such a transformation does not feature retroreflective behavior when a reflective surface is placed at the flattened surface, as shown in Figure \ref{fig:hemispherical}.
In Figure \ref{fig:hemispherical}(b), we see the focal point has been shifted significantly off of the flattened surface, in agreement with Ref. \cite{xu2022hemispherical}.
For the far field results in Figure \ref{fig:hemispherical}(d), note that for $15\degree$ and $45\degree$ plane wave incidence angles, the RCS peaks at $-15\degree$ and $-45\degree$, indicating behavior closer to specular reflection.
Additionally, the RCS at $0\degree$ is significantly smaller than expected for a flat plate of equivalent aperture.
Thus, the hemispherical Luneburg lens does not achieve retroreflection (which we note that it was not designed for).
It may be possible to achieve retroreflection with the hemispherical Luneburg lens with an optimized, non-planar reflective surface, although that is beyond the scope of this work.

\section{\break Random masks distribution}

\begin{figure}[htbp]
\includegraphics[width=\columnwidth]{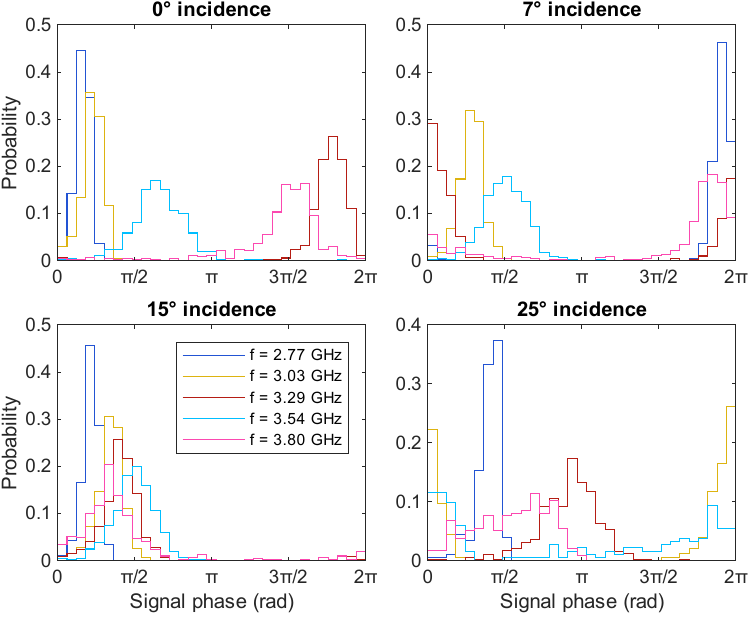}
\caption{Histograms of backscatter phase of random spatial masks applied to the MBLL. Masks are filtered out by magnitude, discarding masks that differ from the maximum magnitude by 10dB.}
\label{fig:random_w_angles}
\end{figure}

As mentioned in Section \ref{sec:multiplex}, the analysis for multiplexed and secure communications relies on post-processing of randomly sampled spatial masks due to experimental limitations.
Figure \ref{fig:random_w_angles} plots the phases of the randomly sampled spatial masks, where data that differ from the maximum magnitude by over 10dB are filtered out. 
For most cases, the phases of the backscatter signal are narrowly distributed within a small range. Thus, with a more sophisticated experimental setup and the application of optimization algorithms inside the experimental loop will likely be able to deliver significantly improved results compared to what has been presented here.

\section*{Acknowledgment}
The authors thank Luke Newton, Matt van Niekerk, Robert Duggan, and Ben Smith for their insights and help.

\bibliographystyle{unsrt}
\bibliography{arxiv}

\begin{IEEEbiography}[{\includegraphics[width=1in,height=1in,clip,keepaspectratio]{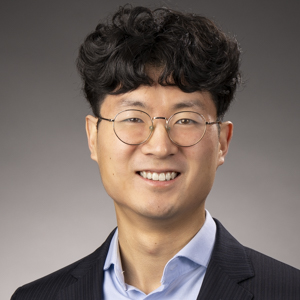}}]
{Samuel Kim} received the A.B. degree in physics from Harvard University, Cambridge, MA in 2015 and the Ph.D. degree in electrical engineering and computer science from the Massachusetts Institute of Technology in 2023.

His graduate work focused on the intersection of physics and machine learning, developing algorithms for scientific discovery, Bayesian optimization, and photonics.
In 2021, he co-founded Kyber Photonics, a spinoff from his research on silicon photonics devices for optical beam steering.
He is currently a research scientist at the Johns Hopkins Applied Physics Laboratory where he focuses on advanced optimization algorithms and computational electromagnetics with applications in metamaterials and integrated photonics. 

Dr. Kim was a recipient of the National Defense Science and Engineering Graduate (NDSEG) Fellowship in 2019.
\end{IEEEbiography}

\begin{IEEEbiography}[{\includegraphics[width=1in,height=1in,clip,keepaspectratio]{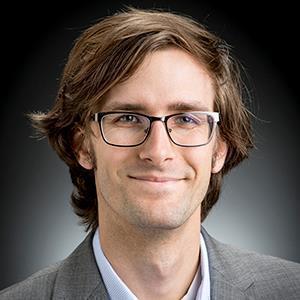}}]{Timothy Sleasman}
(Member, IEEE) received the B.S. degree in mathematics and physics from Boston College, Chestnut Hill, MA, USA, in 2013 and the Ph.D. degree from the Department of Electrical and Computer Engineering, Duke University, Durham, NC, USA, in 2018.

From 2013 to 2018, he was with the Center for Metamaterials and Integrated Plasmonics, Duke University. He is currently a Senior Researcher with the John Hopkins University Applied Physics Lab, Laurel, MD, USA. His current research interests include computational imaging, dynamically tunable metasurfaces, and novel platforms for generating tailored electromagnetic wavefronts.
\end{IEEEbiography}

\begin{IEEEbiography}[{\includegraphics[width=1in,height=1in,clip,keepaspectratio]{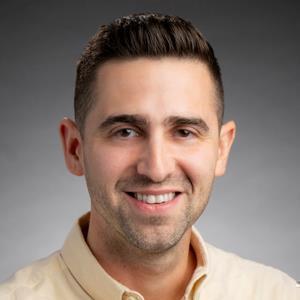}}]{Avrami Rakovsky} is a Mechanical Engineer at the Johns Hopkins University Applied Physics Laboratory (JHU/APL). Avrami received a B.S. in Industrial Design from the New Jersey Institute of Technology in 2012, providing a foundation in human-centered design principles. Building upon this, he earned a M.S. in Mechanical Engineering from Johns Hopkins University, in 2022. 

Driven by a passion for solving real-world challenges, Avrami specializes in translating design concepts into tangible, functional products. His expertise encompasses product design, material development, and the application of specialized software, utilizing a multidisciplinary approach to innovation. He focuses on advanced materials and fabrication methods, with his current research exploring the dynamic relationship between product design and advanced material development.
\end{IEEEbiography}

\begin{IEEEbiography}[{\includegraphics[width=1in,height=1in,clip,keepaspectratio]{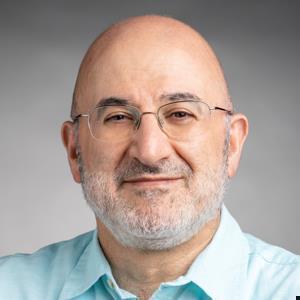}}]{Ra'id Awadallah} 
received his Ph.D. in Electrical Engineering from Virginia Tech in 1998. He joined Johns Hopkins University Applied Physics Laboratory in the same year and he is currently a member of the principal professional staff.  Over the last 27 years, he has led a team of researchers developing efficient numerical models for tropospheric propagation, electromagnetic scattering from randomly rough surfaces, radar cross-section of complex targets, pulsed propagation in complex urban structures, and design and characterization of thin-film metamaterials for various applications. Dr. Awadallah is a member of the IEEE and Commission F or URSI. He is also a lecturer at the Johns Hopkins University Engineering Program for professionals where he teaches courses on basic and applied electromagnetism.
\end{IEEEbiography}

\begin{IEEEbiography}[{\includegraphics[width=1in,height=1in,clip,keepaspectratio]{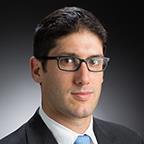}}]{David Shrekenhamer}
is a Staff Scientist at the Johns Hopkins University Applied Physics Laboratory (JHU/APL) and has a secondary research scientist position in the Department of Electrical and Computer Engineering at Johns Hopkins University. Dr. Shrekenhamer received a B.S. degree in physics from University of California San Diego, La Jolla, CA in 2006 and the Ph.D. degree in physics and from Boston College, Chestnut Hill, MA in 2013. 

His research interests span numerous areas of electromagnetics, and he has publications and expertise developing concepts that utilize metamaterials for use from DC to light. At APL, he specializes in leading interdisciplinary teams designing custom electromagnetic solutions for a range of complex problem spaces e.g., communications, signature science, sensing and imaging. While his work is largely focused on electromagnetic-based problems, he also has substantial expertise in developing novel materials (e.g. optical phase change materials), fabrication techniques (e.g. nanofabrication and incorporation within novel host materials and form factors), electronic circuitry and signal processing, and incorporating and accounting for multi-physics effects such as electromagnetic heating, structural heating, and more exotic behavior like nonlinear effects, chirality, and bi-anisotropy.
\end{IEEEbiography}

\EOD

\end{document}